\documentclass[twocolumn,showpacs,amsmath,amssymb,pra]{revtex4}

\usepackage{graphicx}   % Include figure files
\usepackage{dcolumn}    % Align table columns on decimal point
\usepackage{bm}         % bold math

\begin{document}

\title{Perturbative calculation of critical exponents \\
       for the Bose-Hubbard model}

\author{Dennis Hinrichs$^1$}
\author{Axel Pelster$^{2,3}$}
\author{Martin Holthaus$^1$}

\affiliation{
	$^1$Institut f\"ur Physik, Carl von Ossietzky Universit\"at,
	D - 26111 Oldenburg, Germany}
\affiliation{
	$^2$Hanse-Wissenschaftskolleg, Lehmkuhlenbusch 4, 
	D - 27753 Delmenhorst, Germany}
\affiliation{
	$^3$Fachbereich Physik und Forschungszentrum OPTIMAS,
	TU Kaiserslautern, D - 67663 Kaiserslautern, Germany}

\date{March 14, 2013}

\begin{abstract}
We develop a strategy for calculating critical exponents for the Mott
insulator-to-superfluid transition shown by the Bose-Hubbard model. Our
approach is based on the field-theoretic concept of the effective potential,
which provides a natural extension of the Landau theory of phase transitions
to quantum critical phenomena. The coefficients of the Landau expansion of 
that effective potential are obtained by high-order perturbation theory. 
We counteract the divergency of the weak-coupling perturbation series by 
including the seldom considered Landau coefficient $a_6$ into our analysis.
Our preliminary results indicate that the critical exponents for both the 
condensate density and the superfluid density, as derived from the 
two-dimensional Bose-Hubbard model, deviate by less than  $1\%$ from the best 
known estimates computed so far for the three-dimensional $XY$ universality 
class. 
\end{abstract}


% 05.30.Jp   Boson systems
% 05.30.Rt   Quantum phase transitions
% 64.60.F-   Equilibrium properties near critical points, critical exponents

\maketitle

\section{Introduction}
\label{sec:1}

The universality of phase transitions is one of the most important concepts
in the theoretical description of critical 
phenomena~\cite{ZinnJustin02,BinneyEtAl98,KleinertSF01}. It implies that 
continuous phase transitions fall into universality classes determined by 
only a few gross properties characterizing the respective system, namely,
the number of components of the order parameter and their symmetry, the
dimensionality of space, and the range of interaction. Renormalization-group 
(RG) theory then predicts that, e.g., critical exponents are identical for 
all systems within a given such class. For instance, the lambda transition 
undergone by liquid $^4$He at the temperature of $2.17$~Kelvin is the 
primary example of the three-dimensional $XY$ universality class, that is, 
the class with a two-dimensional (or complex) order parameter with $O(2)$ 
symmetry in three spatial dimensions, and with short-range interactions. Thus, 
the critical exponent describing the specific-heat singularity at the 
lambda point, which was found to be $\alpha = -0.0127 \pm 0.0003$ 
in an elaborate zero-gravity experiment~\cite{LipaEtAl03}, 
should coincide with the corresponding exponent predicted by $\Phi^4$~theory. 
Indeed, a seven-loop expansion in three dimensions has resulted in the value 
$\alpha = -0.01126 \pm 0.0010$~\cite{Kleinert00}, while 
$\alpha = -0.0146 \pm 0.0008$ has been obtained by combining Monte Carlo
simulations based on finite-size scaling methods with high-temperature
expansions~\cite{CampostriniEtAl01}. Evidently these two theoretical estimates
bracket the experimental value, but do not agree with it, nor with themselves, 
within the margins of uncertainty stated. Thus, this core test of RG theory is 
not fully conclusive yet; if one accepts the experimental result there still
is a need to improve the theoretical calculations.      

In this situation it may be of interest to observe that the notion of 
universality also includes quantum phase transitions, that is, transitions 
which occur at zero temperature upon variation of a parameter of the 
system under consideration, being triggered by quantum rather than thermal 
fluctuations~\cite{Sachdev01}. In particular, the Mott insulator-to-superfluid 
transition exhibited by the Bose-Hubbard model on a $d$-dimensional cubic 
lattice falls into the universality class of the $(d+1)$-dimensional 
$XY$ model at special multicritical points with particle-hole 
symmetry~\cite{FisherEtAl89}, implying that the critical exponents provided 
by the two-dimensional (2D) Bose-Hubbard model should agree with those 
of the lambda transition. Now that this 2D Bose-Hubbard model has been 
emulated with ultracold $^{87}$Rb atoms loaded into stacks of planar optical 
lattices~\cite{KoehlEtAl05,SpielmanEtAl07}, and even the condensate fraction 
of such a Bose gas in a 2D lattice has been measured across the Mott 
insulator-to-superfluid transition~\cite{SpielmanEtAl08}, future precision
experiments on this system might enable one to accurately determine the 
corresponding critical exponents, and thus to provide a further nontrivial test 
of universality. Indeed, the exploration of critical behavior with ultracold 
dilute quantum gases has already been taken up by Donner {\em et al.\/}, 
who have measured the critical exponent of the correlation length for a
harmonically trapped, weakly interacting 3D Bose gas, albeit with a still 
comparatively large error bar~\cite{DonnerEtAl07}.

On the theoretical side, the archetypal Bose-Hubbard model lends itself to 
alternative computational schemes. Only recently, Ran\c{c}on and Dupuis have 
presented a detailed RG approach to this model, taking into account both local 
and long-distance fluctuations~\cite{RanconDupuis11}. Somewhat alarmingly, the 
numerical value of the critical exponent for the correlation length of the 
2D system derived from that study amounts to $\nu = 0.699$, differing quite 
substantially from the value $\nu = 0.67155 \pm 0.00027$ previously reported 
by Campostrini {\em et al.\/}~\cite{CampostriniEtAl01}. This finding appears 
to put universality into question, and hence calls for further independent 
calculations. In the present paper we establish a ``hands-on'' approach to the 
critical exponents of the Bose-Hubbard model, based on the field-theoretic 
concept of the effective potential~\cite{ZinnJustin02,KleinertSF01}, which 
opens a natural bridge to Landau's theory of phase 
transitions~\cite{SantosPelster09,BradlynEtAl09}. We focus on the exponent 
$\beta_{\rm c}$ for the condensate density, and on the exponent $\zeta$ for 
the superfluid density, from which one can deduce all other critical exponents 
by exploiting (hyper-)scaling relations~\cite{FisherEtAl73,RudnickJasnow77}. 
We proceed as follows: In Sec.~\ref{sec:2} we retrace the basic 
steps required for deriving the Landau expansion of the effective
potential~\cite{SantosPelster09,BradlynEtAl09}, and explain how this 
expansion is employed for computing both the condensate and the superfluid 
density. In Sec.~\ref{sec:3} we recapitulate the idea of the process chain 
approach~\cite{Eckardt09}, which yields perturbative approximants to the 
individal Landau coefficients. The results obtained by evaluating the 
perturbation series numerically to high orders in the hopping strength are 
then discussed at length in Sec.~\ref{sec:4}. Here we encounter a vexing
problem, namely, the divergency of weak-coupling perturbation theory. In
principle, this calls for a systematic resummation procedure for deducing
the ``true'', regular  Landau coefficients from their diverging polynomial
approximants. Nonetheless, here we show that even {\em without\/} such a 
procedure, but by explicitly including the seldom considered Landau coefficient 
$a_6$ into the analysis, one is able to extract critical exponents for the 
2D~Bose-Hubbard system which agree to better than $1\%$ with those 
computed for the lambda transition~\cite{CampostriniEtAl01}, thus providing 
fair evidence in favor of universality. Our {\em ad hoc\/} procedure still 
requires formal justification and hence should be regarded as preliminary, 
but quite similar results are obtained by applying variational perturbation 
theory~\cite{HinrichsEtAl13}. Some conclusions are drawn in the final 
Sec.~\ref{sec:5}.

\section{The method of the effective potential}
\label{sec:2}

The pure Bose-Hubbard model describes Bose particles on a lattice which are
allowed to tunnel between neighboring lattice sites, while repelling each other
when occupying the same site. In terms of operators $\widehat{b}_i^{\dagger}$ 
and $\widehat{b}_i^{\phantom \dagger}$ which encode the creation and 
annihilation of a Bose particle at the $i$th site and thus obey the commutation
relation
\begin{equation}
	[ \widehat{b}_i^{\phantom \dagger}, \widehat{b}_j^{\dagger} ]
	= \delta_{ij} \; ,
\end{equation} 
it is defined by the grand-canonical Hamiltonian~\cite{FisherEtAl89} 
\begin{equation}
	\widehat{H}_{\rm BH} = \widehat{H}_0 + \widehat{H}_{\rm tun} \; ,
\label{eq:HBH}
\end{equation}	
where the site-diagonal part 
\begin{equation}
	\widehat{H}_0 = \frac{1}{2} \sum_{i} \widehat{n}_i(\widehat{n}_i-1) 
	- \mu/U \sum_{i} \widehat{n}_i
\label{eq:SDC}
\end{equation} 
models the on-site repulsion, and fixes the total particle number through the 
adjustment of the chemical potential $\mu$. Here, 
\begin{equation}
	\widehat{n}_i = 
	\widehat{b}_i^{\dagger}\widehat{b}_i^{\phantom \dagger}
\end{equation}
counts the number of particles at the $i$th site, and $U$ is the repulsion
energy contributed by any pair of particles sitting on a common site. We are
using this energy $U$ as scale of reference for writing the Hamiltonian in 
dimensionless form. On the other hand, denoting the energy associated with a 
hopping event by $J$, nearest-neighbor tunneling of the particles is described 
by   		 
\begin{equation}
	\widehat{H}_{\rm tun} = - J/U \sum_{\langle i,j \rangle} \, 
	\widehat{b}_i^{\dagger}\widehat{b}_j^{\phantom \dagger}  \; , 
\label{eq:HTN}
\end{equation}
with the angular brackets under the sum indicating that $i$ and $j$ are 
restricted to pairs of adjacent sites. As is well known, the 
particle-delocalizing tendency of $\widehat{H}_{\rm tun}$ counteracts the
localizing tendency of the repulsive interaction, so that the system 
exhibits a transition from a Mott insulator to a superfluid when the 
control parameter $J/U$ is enhanced gradually, while the scaled chemical 
potential $\mu/U$ is kept constant~\cite{Sachdev01,FisherEtAl89}.   

In order to map out this quantum phase transition, one studies the system's
reaction to the attempt to couple particles into or out of the lattice
through spatially homogeneous sources and drains, as expressed by the
extended Hamiltonian
\begin{equation}
	\widehat{H} = \widehat{H}_{\rm BH} + \widehat{H}_{\rm s-d} \; ,
\label{eq:EXH}
\end{equation}
where
\begin{equation}
	\widehat{H}_{\rm s-d} = \sum_i \left( \eta \, \widehat{b}_i^{\dagger}
	+ \eta^* \widehat{b}_i^{\phantom \dagger} \right) \; . 
\label{eq:HSD}
\end{equation}
Formally, this step corresponds to explicitly breaking the global 
particle-number conservation built into $\widehat{H}_{\rm BH}$, the intuitive 
idea being that the system should resist this attempt for sufficiently small 
source strength $\eta$ when being in a Mott insulator state, but show some 
response for any nonzero $\eta$ in the superfluid state.  

Restricting ourselves to zero temperature, the free energy $\mathcal{F}$ of 
the extended system is given by the ground-state expectation value of its
Hamiltonian, 
\begin{equation}
	\mathcal{F}(J/U, \mu/U, \eta, \eta^*) 
	= \langle \widehat{H} \rangle \; .
\label{eq:GEV}	
\end{equation}	
Assuming the lattice to consist of $M$ sites (while stipulating that the 
thermodynamic limit $M \to \infty$ be taken eventually), we expand this free 
energy in the form
\begin{eqnarray}
\label{eq:EFE}
	& & \mathcal{F}(J/U, \mu/U, \eta, \eta^*) \\
	& = &
	M \left(f_0(J/U, \mu/U) + \sum_{k=1}^{\infty} 
	c_{2k}(J/U, \mu/U) \, |\eta|^{2k} \right) \; ,
	\nonumber	
\end{eqnarray}
so that $f_0$ denotes the free energy per site of the original 
system~(\ref{eq:HBH}). The fact that $\mathcal{F}$ is expressed here in powers
of $|\eta|^2$, rather than of $\eta$ and $\eta^*$ individually, is understood 
from the perturbative viewpoint adopted in the following section: If one regards
the creation and annihilation operations implementing these sources and drains 
as individual perturbation events, it is obvious that only processes with an 
equal number of creation and annihilation events, and hence terms with equal 
powers of $\eta$ and $\eta^*$, can contribute to the expectation 
value~(\ref{eq:GEV}).

Following the guiding insight that the response of the system to the sources
or drains, and hence the change of $\mathcal{F}$ with $\eta$ or $\eta^*$,
should reveal its state, it is only natural to consider the intensive
quantities
\begin{equation}
	\psi = \frac{1}{M}\frac{\partial\mathcal{F}}{\partial \eta^*}
	= \langle \widehat{b}_i^{\phantom \dagger} \rangle
	\quad , \quad
	\psi^* = \frac{1}{M}\frac{\partial\mathcal{F}}{\partial \eta}
	= \langle \widehat{b}_i^{\dagger} \rangle \; .	
\label{eq:DOP}
\end{equation}
The respective first equalities in these two relations are nothing but
definitions of $\psi$ and $\psi^*$, whereas the respective second equalities
follow immediately from the Hellmann-Feynman theorem~\cite{Feynman39,Fitts99}. 
Of course, this is the standard way in field theory to introduce the order
parameter~\cite{ZinnJustin02,KleinertSF01}.   

The decisive step now is to take $\psi$ and $\psi^*$ as new independent 
variables. This is accomplished by performing a Legendre transformation of 
$\mathcal{F}$, thus constructing the effective potential~\cite{SantosPelster09}    
\begin{equation}
	\Gamma(J/U, \mu/U, \psi, \psi^*) = \mathcal{F}
	- M(\eta^*\psi + \eta\psi^*) \; ,
\label{eq:DEP}
\end{equation}
where the old variables $\eta$ and $\eta^*$ have to be expressed in terms of 
$\psi$ and $\psi^*$. To this end, combining the definition~(\ref{eq:DOP}) 
with the expansion~(\ref{eq:EFE}) gives
\begin{equation}
	\psi = \eta \Big[ c_2 + 2 c_4 |\eta|^2 + 3 c_6 |\eta|^4
	+ \mathcal{O}(|\eta|^6) \Big]
\end{equation}
and its complex conjugate, which then yields 
\begin{equation}
	\eta = \psi \left[ \frac{1}{c_2} - \frac{2c_4}{c_2^4} |\psi|^2
	+ \left( \frac{12c_4^2}{c_2^7} - \frac{3c_6}{c_2^6}\right) |\psi|^4
	+ \mathcal{O}(|\psi|^6) \right]
\end{equation}
upon inversion. Inserting, one obtains the effective potential~(\ref{eq:DEP}) 
as a series in powers of $|\psi|^2$:	
\begin{equation}
	\frac{1}{M}\Gamma = 
	f_0 + a_2 |\psi|^2 + a_4 |\psi|^4 + a_6 |\psi|^6
	+ \mathcal{O}(|\psi|^8) 
\label{eq:EFP}
\end{equation} 
with coefficients
\begin{eqnarray}
	a_2 & = & -\frac{1}{c_2} \; ,
\nonumber	\\
	a_4 & = & \frac{c_4}{c_2^4} \; ,
\nonumber	\\
	a_6 & = & \frac{c_6}{c_2^6} - \frac{4c_4^2}{c_2^7} \; ,
\label{eq:COP}
\end{eqnarray}
having suppressed their dependence on $J/U$ and $\mu/U$.			
	
So far, these elementary considerations still refer to the extended 
system~(\ref{eq:EXH}), from which the original Bose-Hubbard model~(\ref{eq:HBH})
is recovered by equating $\eta = \eta^* = 0$. By construction, $\eta$ and 
$\psi^*$ on the one hand, and $\eta^*$ and $\psi$ on the other, each constitute
a Legendre-conjugated pair~\cite{Arnold89}, so that one also has  
\begin{equation}
	\frac{1}{M}\frac{\partial\Gamma}{\partial \psi^*} = - \eta
	\quad , \quad
	\frac{1}{M}\frac{\partial\Gamma}{\partial \psi} = - \eta^* \; . 	
\label{eq:LCP}
\end{equation}
This is what finally explains why $\Gamma$ has suggestively been named 
``effective potential'': Setting $\eta = \eta^* = 0$ in these 
equations~(\ref{eq:LCP}) means that the order parameter $\psi_0$ describing 
the actual Bose-Hubbard system~(\ref{eq:HBH}) is determined by finding a
stationary point of $\Gamma$, in the same manner as a mechanical equilibrium 
is determined by a stationary point of some given mechanical potential, with 
stable equilibria corresponding to minima.  

Now we can virtually copy the Landau theory of phase transitions: Assuming 
$a_4$ and $a_6$ to be positive, and neglecting higher-order terms of the 
effective potential~(\ref{eq:EFP}), the minimum of $\Gamma$ is found at 
$\psi_0 = 0$ as long as $a_2 > 0$, which indicates the Mott insulator phase. 
In contrast, when $a_2 < 0$ the order parameter takes on a nonzero value, 
signaling the presence of the superfluid phase. Since $| \psi_0 |^2$ then is 
to be identified with the condensate density $\varrho_{\rm c}$, one has
\begin{equation}
	\varrho_{\rm c} = |  \psi_0 |^2 = \frac{1}{3 a_6}
	\left( -a_4 + \sqrt{a_4^2 - 3 a_2 a_6} \right)
\label{eq:CDN}
\end{equation}
when $a_2 < 0$. Thus, knowledge of solely the coefficient $a_2(J/U,\mu/U)$ 
already enables one to locate the phase boundary by means of the condition 
$a_2 = 0$~\cite{SantosPelster09}; if one possesses still more information on 
the effective potential, in the guise of the higher coeffients $a_4$ and 
$a_6$, say, one can even monitor the appearance of the order parameter when 
that boundary is crossed, and hence determine the critical exponent 
$\beta_{\rm c}$ of the condensate density.  

For computing also the superfluid density $\varrho_{\rm s}$ and its critcial
exponent $\zeta$, we recall that if
\begin{equation}
	\chi_0(\mathbf{x}) = \exp\big({\rm i}\varphi(\mathbf{x})\big) \, 
	|\chi_0(\mathbf{x})| 
\end{equation}	 
is a single-particle state macroscopically occupied by Bose particles of mass
$m$, the superfluid velocity $\mathbf{v}_{\rm s}(\mathbf x)$ is defined by the
relation~\cite{Leggett99}
\begin{equation}
	\mathbf{v}_{\rm s}(\mathbf x) 
	= \frac{\hbar}{m} \nabla \varphi(\mathbf x) \; .
\end{equation}	
Dealing with a $d$-dimensional hypercubic lattice, it is convenient to adopt 
the particular choice
\begin{equation}
	\varphi(\mathbf{x}) = \theta \, \mathbf{e} \cdot \mathbf{x} / L \; ,
\end{equation}
where $\mathbf{e}$ is a unit vector in the direction of an arbitrary lattice
axis, all of which are equivalent. This means that the phase progresses by 
the twist angle $\theta$ on each path of length $L$ parallel to $\mathbf{e}$.
The twist is imposed on the many-body wave function $\Psi$ by 
requiring~\cite{ShastrySutherland90,RothBurnett03} 
\begin{equation}
	\Psi( \ldots, \mathbf{x}_j + L \mathbf{e}, \ldots) 
	= {\rm e}^{{\rm i}\theta} \Psi( \ldots, \mathbf{x}_j, \ldots)
\end{equation}	
for each particle (labeled here by $j$). Operationally, this is achieved by 
performing the local unitary transformation    
\begin{equation}
	\widehat{b}_i^{\phantom \dagger} \to 
	{\rm e}^{{\rm i}\varphi(\mathbf{x}_i)} \, 
	\widehat{b}_i^{\phantom \dagger}
	\quad , \quad
	\widehat{b}_i^{\dagger} \to 
	{\rm e}^{-{\rm i}\varphi(\mathbf{x}_i)} \, 
	\widehat{b}_i^{\dagger} \; ,	
\label{eq:PHT}
\end{equation}
where $\mathbf{x}_i$ is the position of the lattice site No.~$i$; in this 
way, the boundary conditions are shifted onto the Hamiltonian. Now let 
$\mathcal{F}(\theta)$ be the free energy~(\ref{eq:GEV}) as belonging to 
the ``twisted'' Hamiltonian which gives rise to superfluid flow, denote the 
number of lattice sites inside the hypercube $L^d$ by $M$, and specify 
$\varrho_{\rm s}$ as the number of superfluid particles per lattice site. If 
the particles were free, this would imply   
\begin{eqnarray}
	U \Big[ \mathcal{F}(\theta) - \mathcal{F}(0) \Big]
	& = & M \varrho_{\rm s} \frac{m}{2} \mathbf{v}_{\rm s}^2
\nonumber \\ & = &	
	M \varrho_{\rm s} \frac{m}{2} \left(\frac{\hbar}{m}\right)^2
	\left(\frac{\theta}{L}\right)^2 \; .
\label{eq:SKE}	
\end{eqnarray}	
But since the single-particle dispersion relation actually reads 
\begin{equation}
	E(\mathbf{k}) = -2J \sum_{j=1}^d \cos(k_j a) \; ,
\end{equation}	
where $a$ is the lattice constant, one has to replace the factor 
$\hbar^2/(2m)$ in Eq.~(\ref{eq:SKE}) by $Ja^2$. Moreover, by virtue of the 
geometrical properties of the Legendre transformation~\cite{Arnold89} the
free energy equals the effective potential when the latter is evaluated at 
its mimimum $\psi_0$~\cite{BradlynEtAl09}. Taken together, this gives 
\begin{equation}
	U \Big[\Gamma(\theta)|_{\psi_0} - \Gamma(0)|_{\psi_0} \Big] = 
	M \varrho_{\rm s} J a^2 \left(\frac{\theta}{L}\right)^2  
\end{equation}
for sufficiently small $\theta/L$. Measuring lengths in multiples of the 
lattice constant and hence writing $\ell = L/a$, this finally leads to
\begin{equation}
	\varrho_{\rm s} = \lim_{\theta \to 0} 
	\frac{1}{M J/U}\left(\frac{\ell}{\theta}\right)^2
	\Big[\Gamma(\theta)|_{\psi_0} - \Gamma(0)|_{\psi_0} \Big] \; . 
\label{eq:RSF}
\end{equation}	
This expression is closely related to the helicity modulus introduced by
Fisher {\em et al.}~\cite{FisherEtAl73}, emphasizing that the superfluid 
density quantifies the rigidity of the system under the imposed twist. Thus, 
sufficient knowledge of the effective potential, both with and without such 
a twist, enables one to monitor the emergence of $\varrho_{\rm s}$ when the 
phase boundary is crossed upon varying $J/U$, and thereby to determine its 
critical exponent~$\zeta$.

\section{The process-chain approach to computing the effective potential}
\label{sec:3}

The main computational task now consists in the calculation of the expansion 
coefficients $a_{2k}$ of the effective potential~(\ref{eq:EFP}), which, 
according to Eq.~(\ref{eq:COP}), are given in terms of the coefficients 
$c_{2k}$ introduced in the expansion~(\ref{eq:EFE}) of the free energy, 
either without or including an additional phase twist~(\ref{eq:PHT}). 
We obtain these coefficients by means of the process-chain approach devised 
by Eckardt~\cite{Eckardt09}, which is based on a formulation of the 
perturbation series going back to the Japanese mathe\-matician Tosio 
Kato~\cite{Kato49,Messiah81}: Consider a Hamiltonian $\widehat{H}_0$ with 
a nondegenerate eigenstate  $| m \rangle$ and corresponding eigenvalue 
$E_m^{(0)}$ which is subjected to some suitable perturbation $\widehat{V}$, 
such that the total Hamiltonian becomes 
$\widehat{H} = \widehat{H}_0 + \widehat{V}$. Then the $n$th-order contribution 
$E_m^{(n)}$ to the perturbation series
\begin{equation}	
	E_m = E_m^{(0)} + \sum_{n=1}^{\infty} E_m^{(n)}
\label{eq:SER}
\end{equation}	
for the eigenvalue $E_m$ of $\widehat{H}$ which evolves from $E_m^{(0)}$
upon turning on the perturbation can be written in the non-recursive
form~\cite{Kato49,Messiah81}	
\begin{equation}	
	E_m^{(n)} = {\rm tr} \left[ \sum_{\Lambda_n} 
	\widehat{S}^{\alpha_1} \widehat{V} \widehat{S}^{\alpha_2} \widehat{V} 
	\widehat{S}^{\alpha_3} \ldots  \widehat{S}^{\alpha_n} \widehat{V} 
	\widehat{S}^{\alpha_{n+1}} \right] \; , 
\label{eq:KTF}
\end{equation}
where the chain operators $\widehat{S}^\alpha$ concatenating the $n$ perturbing 
operators $\widehat{V}$ are given by
\begin{equation}
	\widehat{S}^{\alpha} = \left\{ \begin{array}{ll} 
	-|m \rangle \langle m| & \quad \mbox{for } \alpha = 0 
	\phantom{\Big|}\\
	\displaystyle
	\sum\limits_{i\neq m} 
	\frac{|i \rangle \langle i|}
	     {\left(E_m^{(0)} - E_i^{(0)}\right)^{\alpha}} 
        & \quad \mbox{for } \alpha > 0 
	\end{array} \right. \; , 
\end{equation}
and the sum extends over all sets of $n+1$ nonnegative integers $\alpha_j$ 
which sum up to $n-1$, 
\begin{equation}
	\Lambda_n = \left\{ (\alpha_1, \ldots, \alpha_{n+1}) \Big|
	\sum_{j=1}^{n+1} \alpha_j = n-1 \right\} \; .
\end{equation}
By means of standard manipulations~\cite{Eckardt09,TeichmannEtAl09b}, the 
individual terms arising from Kato's trace formula~(\ref{eq:KTF}) can be cast 
into matrix elements of the form
\begin{equation}       
        \langle m| 
	\widehat{V} \widehat{S}^{\alpha_1} \widehat{V} \widehat{S}^{\alpha_2} 
	\ldots \widehat{S}^{\alpha_{n-1}} \widehat{V} 
	|m \rangle \; ,
\label{eq:KAT}
\end{equation}	
to be multiplied with certain weight factors. These matrix elements allow for
an intuitive interpretation: Starting from the initial state $|m\rangle$, the 
system undergoes a chain of $n$ subsequent perturbation processes before 
finally returning to $|m\rangle$. If there are no selection rules making some
of these matrix elements vanish, their number increases by a factor of more 
than 2 when advancing from $n$ to $n+1$: One faces 10 elements in $5$th order, 
but already 627 for $n = 10$~\cite{Eckardt09,TeichmannEtAl09b}. 
 
\begin{figure}
\centering
\includegraphics[scale=0.22,angle=0]{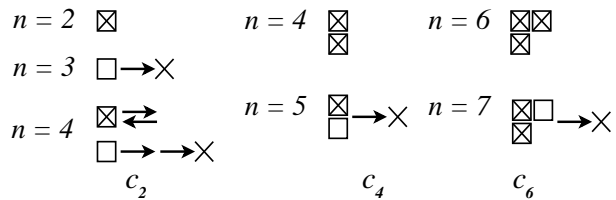}
\caption{Diagrammatic representation of the lowest-order contributions to 
	the quantities $c_2$, $c_4$, and $c_6$. Creation and annihilation 
	processes are symbolized by open boxes and crosses, respectively;
	an arrow denotes a tunneling process between neighboring lattice 
	sites. In general, the coefficients $\gamma_{2k}^{(\nu)}$ introduced
	in the formal expansion~(\ref{eq:FEX}) incorporate all chains with 
	$k$ creation events, $k$ annihilation events, and $\nu = n-2k$
	tunneling events. The determination of all such diagrams, and their
	respective weights, is accomplished by the process-chain approach.}
\label{fig:1}
\end{figure}

In our case, the ``unperturbed'' operator $\widehat{H}_0$ is given by the 
site-diagonal component~(\ref{eq:SDC}) of the Bose-Hubbard Hamiltonian, the 
eigenstates of which are characterized by sharp occupation numbers for each 
lattice site. We consider a Mott state with integer filling factor $g$, that 
is, a state with $g$ particles residing on each site: 
\begin{equation}
	| m \rangle = \prod_i 
	\frac{\left(\widehat{b}_i^{\dagger}\right)^g}
        {\sqrt{g!}} | 0 \rangle \; ,  
\end{equation} 
where $| 0 \rangle$ is the empty-lattice state. In what follows we restrict 
ourselves to $g = 1$, meaning that we have to adjust $\mu/U$ accordingly. 
The perturbation is given by the tunneling Hamiltonian~(\ref{eq:HTN}) 
combined with the sources and drains described by the symmetry-breaking 
extension~(\ref{eq:HSD}), so that  
\begin{equation}
	\widehat{V} = \widehat{H}_{\rm tun} + \widehat{H}_{\rm s-d} \; ,
\end{equation}	
and the goal is to evaluate the perturbation series~(\ref{eq:SER}) for 
$\langle \widehat{H} \rangle = E_m$. Now the representation~(\ref{eq:EFE}) 
tells us that the desired quantities $c_{2k}$ emerge as prefactors of
$|\eta|^{2k}$ in a series expansion of $E_m/M$ with respect to powers of
$|\eta|^2$, and therefore are given by all process chains containing $k$ 
creation operators $\widehat{b}_i^{\dagger}$ and further $k$ annihilation 
operators $\widehat{b}_j^{\phantom \dagger}$. Hence, when considering a 
formal hopping expansion of these functions,  	
\begin{equation}
	c_{2k} = \sum_{\nu=0}^\infty \gamma_{2k}^{(\nu)} (J/U)^\nu \; ,
\label{eq:FEX}
\end{equation}
$n$th-order perturbation theory gives access to the coefficients
$\gamma_{2k}^{(\nu)}$ with $\nu = n-2k$, assuming $n \ge 2k$. By construction, 
these coefficients embody the collection of all process chains with $k$ 
creation and $k$ annihilation events, and $n-2k$ additional hopping events; 
a diagrammatic representation of the lowest-order contributions to $c_2$, 
$c_4$, and $c_6$ is depicted in Fig.~\ref{fig:1}. When mastering this 
process-chain approach in higher orders, the computational bottleneck does not 
lie in the determination of the comparatively few Kato terms~(\ref{eq:KAT}), 
but rather in the fact that for each such term one has to consider all 
permutations of the respective processes~\cite{TeichmannEtAl09b} --- requiring 
us to deal with $12! = 479.001.600$ permutations for $n = 12$, which is the 
maximum order considered in the present paper.

Nonetheless, this process-chain approach can be implemented in a numerically 
efficient manner. So far, we have employed this technique for computing 
accurate phase boundaries for cubic lattices with arbitrary filling 
factors~\cite{TeichmannEtAl09b,TeichmannEtAl09a}, for establishing a scaling 
property of the critical hopping strengths~\cite{TeichmannHinrichs09}, and 
for determining the critical parameters for both triangular and hexa\-gonal 
lattices~\cite{TeichmannEtAl10}. In a more recent study of Bose-Hubbard and
Jaynes-Cummings lattice models the process-chain approach has been judged 
to be extremely powerful~\cite{HeilvdLinden12}; a closely related scheme has 
been utilized successfully for evaluating high-order terms for the fermionic
Hubbard model~\cite{KalinowskiGluza12}. In the following chapter we will report
our preliminary results obtained when applying the perturbative process-chain 
approach to the determination of the effective potential of the Bose-Hubbard 
model, and, in a straightforward further step, to the calculation of critical 
exponents.

\section{Results}
\label{sec:4}

Having gone through the preceding deliberations, the roadmap now is plainly
laid out: The process-chain approach is employed for computing polynomial 
approximations to the coefficients $c_{2k}(J/U,\mu/U)$. These are rearranged to 
provide corresponding approximations to the coefficients $a_{2k}(J/U,\mu/U)$
appearing in the Landau expansion~(\ref{eq:EFP}) of the effective potential,
from which one then obtains the condensate density $\varrho_{\rm c}$ and, 
after inclusion of a phase twist, the superfluid density $\varrho_{\rm s}$.

\begin{figure}
\centering
\includegraphics[scale=0.5,angle=0]{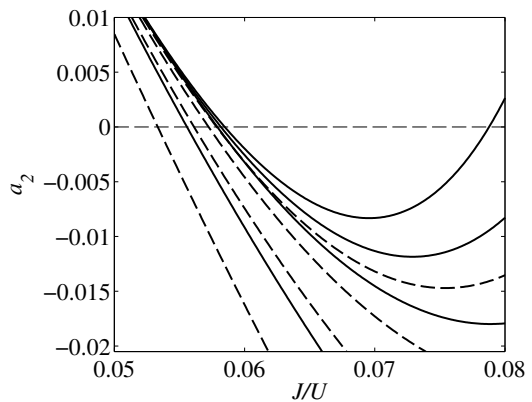}
\caption{Successive perturbational approximants to the Landau coefficient 
	$a_2$ for the 2D Bose-Hubbard model with scaled chemical potential 
	$(\mu/U)_{\rm c} = 0.373$, as corresponding to the tip of the Mott 
	lobe with filling factor $g = 1$. Starting with the leftmost line, 
	and proceeding counter-clockwise, the respective maximum hopping 
	orders $\nu_{\rm m}$ are 3, 2, 5, 7, 4, 9, 6, 8. Here and in the
	following figures~\ref{fig:5}--\ref{fig:7}, full lines refer to even
	and  dashed lines to odd $\nu_m$.}
\label{fig:2}
\end{figure}

\begin{figure}
\centering
\includegraphics[scale=0.5,angle=0]{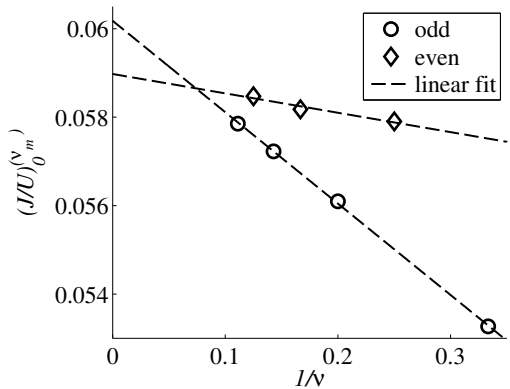}
\caption{Zeros $(J/U)_0^{(\nu_{\rm m})}$ of the approximants to $a_2$
	shown in Fig.~\ref{fig:2}, plotted versus the inverse maximum hopping 
	order. Observe that data points belonging to odd or even $\nu_m$ can 
	be fitted separately to straight lines. The extrapolations of these
	lines to the left margin provide upper and lower bounds on the
	critical scaled hopping strength $(J/U)_{\rm c}$.}	
\label{fig:3}
\end{figure}

Figure~\ref{fig:2} shows results for the coefficient $a_2$ for the 2D
Bose-Hubbard model with fixed chemical potential $(\mu/U)_{\rm c} = 0.373$, 
as corresponding to the border between the Mott insulator and the superfluid 
state with filling factor $g = 1$ (see also Fig.~\ref{fig:4} below). Maximum 
hopping orders taken into account here range from $\nu_m = 2$ to $\nu_m = 9$, 
matching the orders $n = 4$ to $n = 11$ of the perturbation series. The zeros 
of the successive approximants to $a_2$, considered as functions of the scaled 
hopping strength $J/U$, mark the respective estimates $(J/U)_0^{(\nu_{\rm m})}$
of the scaled critical hopping strength $(J/U)_{\rm c}$ for $g = 1$; 
these zeros are plotted in Fig.~\ref{fig:3} over the inverse hopping order. 
Evidently, data points resulting from odd and even $\nu_{\rm m}$ can separately 
be fitted to straight lines; the extra\-polations of these lines for 
$\nu_{\rm m} \to \infty$, or $1/\nu_{\rm m} \to 0$, should contain information 
on the true value of $(J/U)_{\rm c}$. Alternatively one can compute the phase 
boundary by means of the ``ratio-test'' method, which amounts to estimating 
the apparent radius of convergence of the series~(\ref{eq:FEX}) for
$c_2$~\cite{TeichmannEtAl09b,TeichmannEtAl09a}, instead of determining the 
zero of $a_2 = -1/c_2$. Including contributions up to $n = 11$, we find
$(J/U)_{\rm c} \approx 0.05920$ in this manner, suggesting that the two
extra\-polated values inferred from Fig.~\ref{fig:3} serve as upper and lower 
bound on the actual value. If one accepts this hypothesis, the possible error 
of our phase boundary is at most on the order of 2\%. Indeed, this estimate 
is well compatible with the result $(J/U)_{\rm c} = 0.05974(3)$ provided by 
quantum Monte Carlo (QMC) simulations~\cite{CapogrossoSansoneEtAl08}.

\begin{figure}
\centering
\includegraphics[scale=0.5,angle=0]{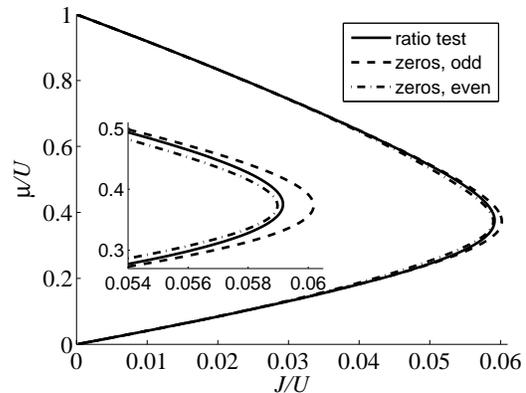}
\caption{Mott lobe with filling factor $g = 1$ for the 2D Bose-Hubbard model,
	computed with the ``ratio-test'' method put forward in 
	Refs.~\cite{TeichmannEtAl09b,TeichmannEtAl09a}. 
	Also shown are the bounds obtained by the procedure sketched in
	Fig.~\ref{fig:3}.}
\label{fig:4}
\end{figure}

Figure~\ref{fig:4} then depicts the entire lowest Mott lobe for the 2D 
Bose-Hubbard model, i.e., the boundary between the Mott insulator state
with $g = 1$ (inside the lobe) and the superfluid state (outside); here the 
result provided by the ratio test is framed by the two bounds determined 
according to the scheme depicted in Fig.~\ref{fig:3}. In order to compute the 
critical exponents of the quantum phase transition, we have to focus on the 
tip of this lobe~\cite{FisherEtAl89}.

\begin{figure}
\centering
\includegraphics[scale=0.5,angle=0]{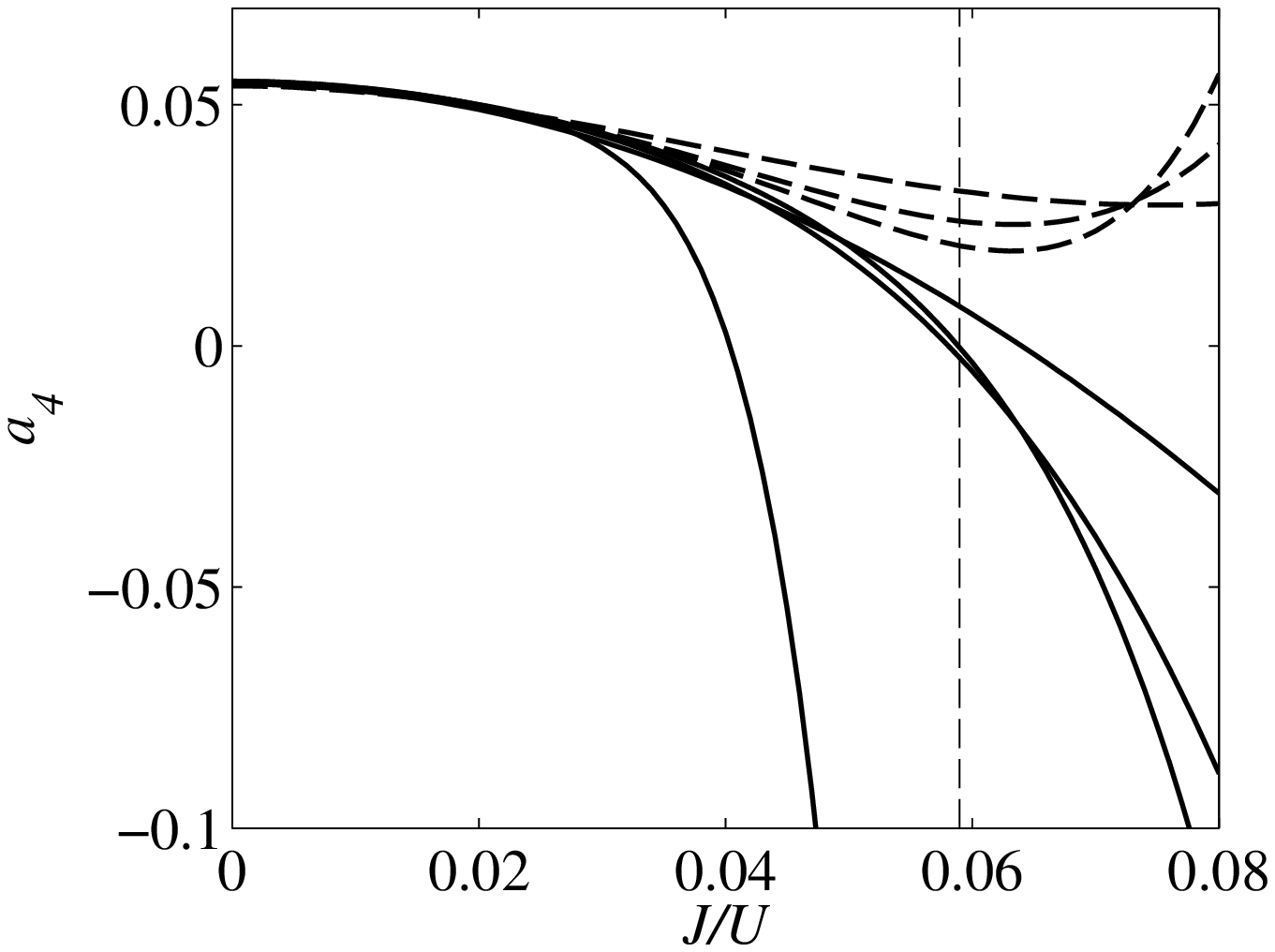}
\includegraphics[scale=0.5,angle=0]{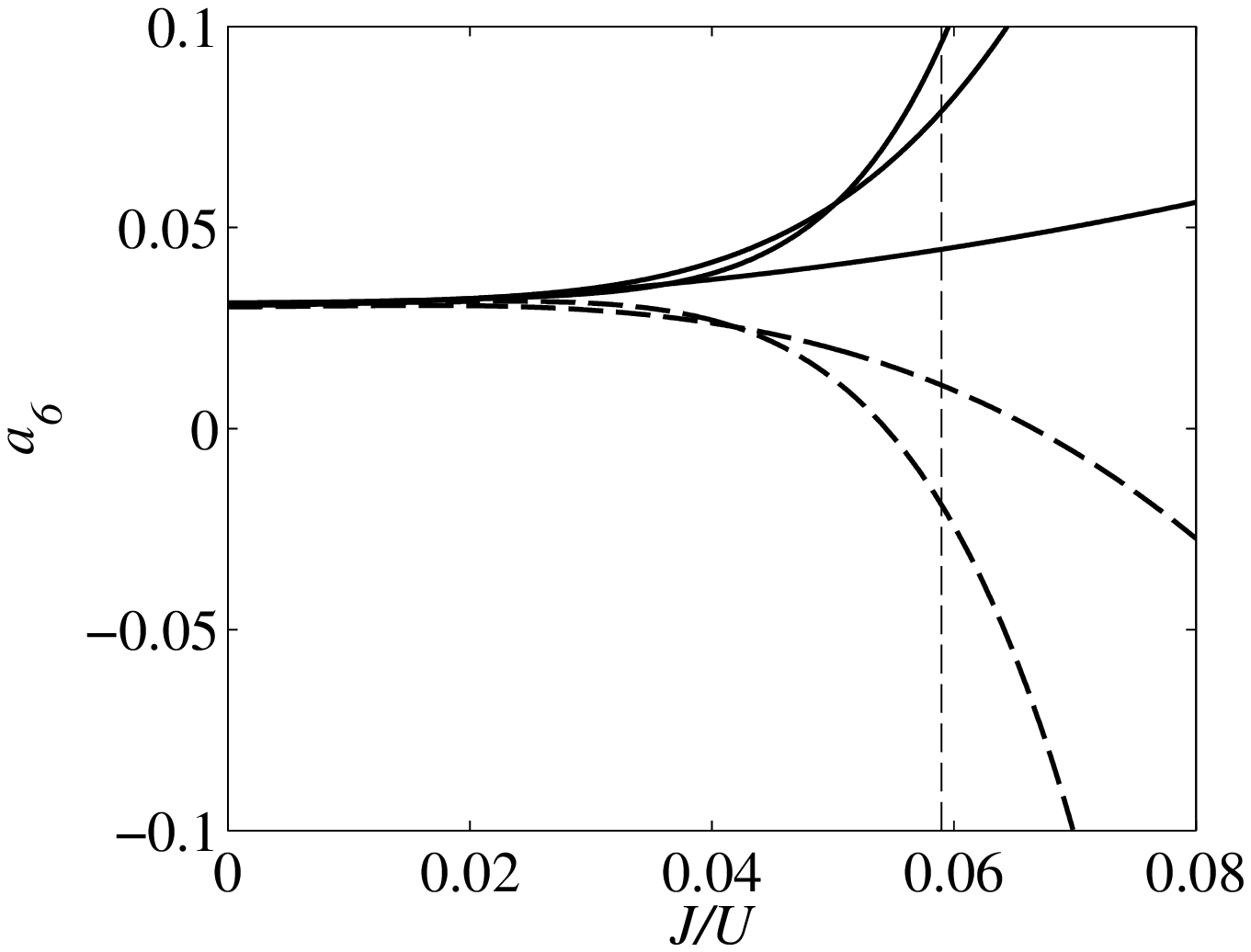}
\caption{Upper panel: Approximants to the coefficient $a_4$ for $d = 2$ 
	and $(\mu/U)_{\rm c} = 0.373$, as in Fig.~\ref{fig:2}. Starting 
	with the leftmost line crossing the lower margin, and proceeding 
	counter-clockwise along the margin, the respective maximum 
	hopping orders $\nu_{\rm m}$ are 8, 6, 4, 2, 3, 5, 7.
	Lower panel: Approximants to the coefficient $a_6$ for $d = 2$ 
	and $(\mu/U)_{\rm c} = 0.373$. Starting with the line crossing 
	the lower margin and proceeding counter-clockwise, the respective 
	maximum hopping orders $\nu_{\rm m}$ are 5, 3, 2, 4, 6. 
	Vertical dashed lines mark $(J/U)_{\rm c}$.} 	
\label{fig:5}
\end{figure}

\begin{figure}
\centering
\includegraphics[scale=0.5,angle=0]{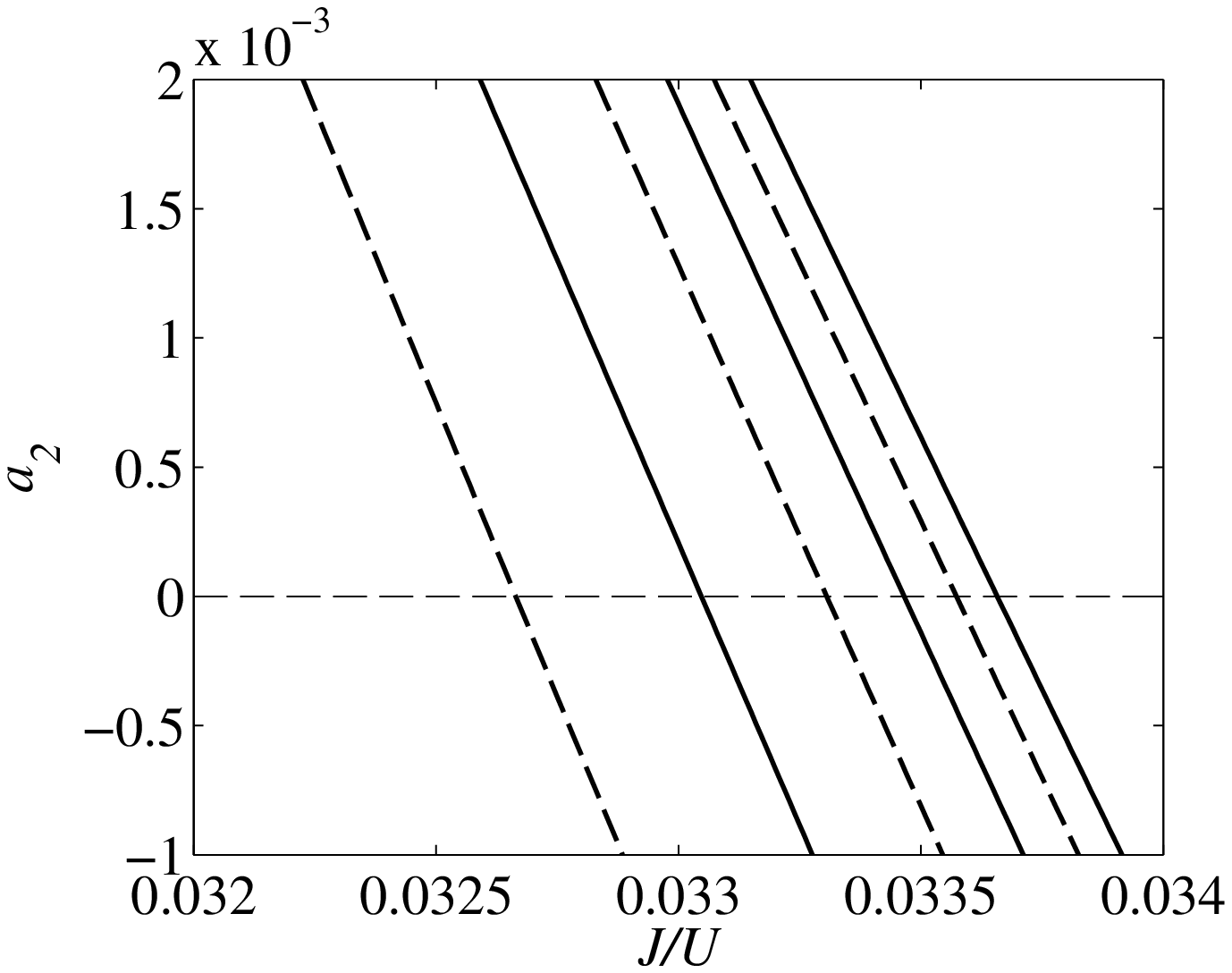}
\includegraphics[scale=0.5,angle=0]{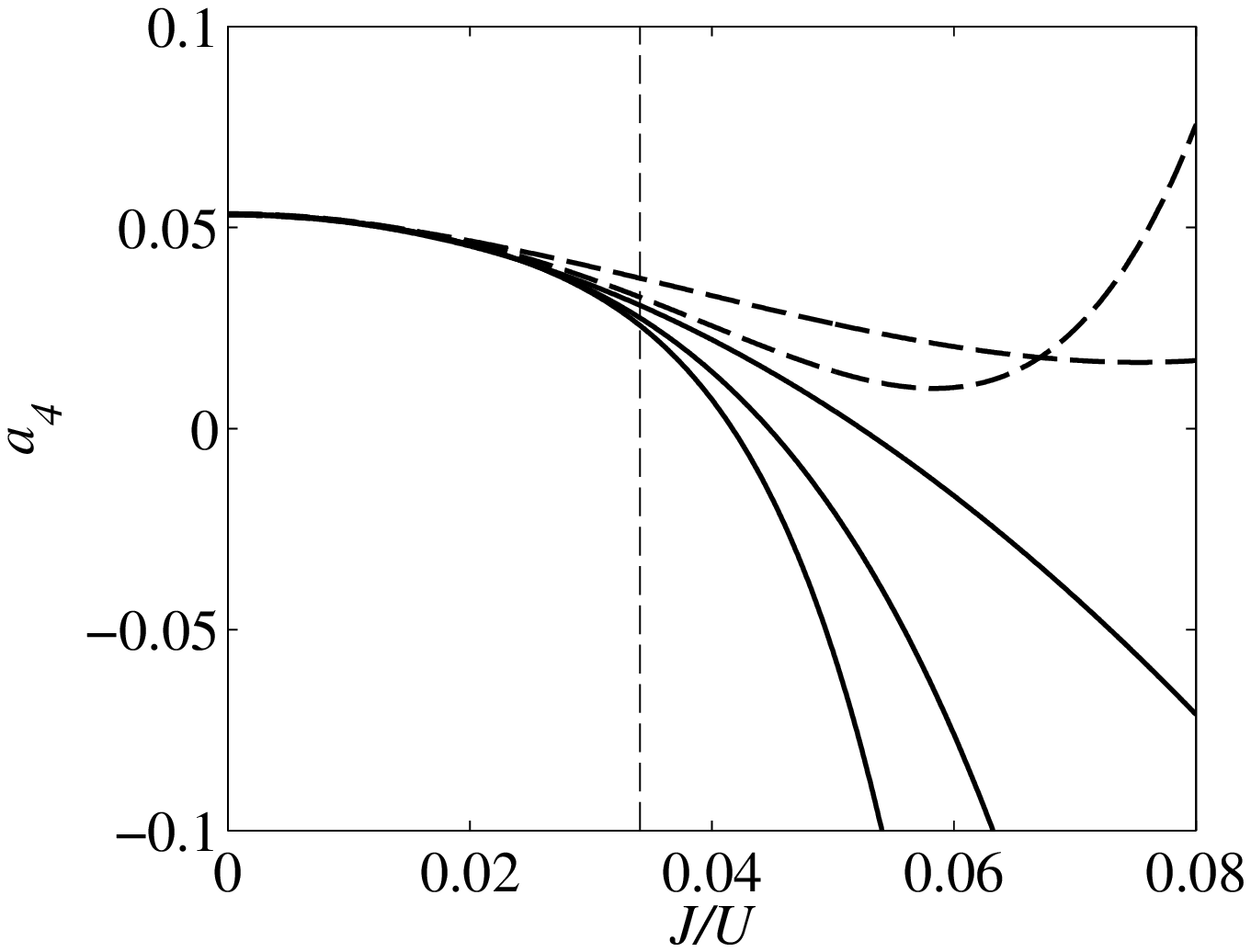}
\includegraphics[scale=0.5,angle=0]{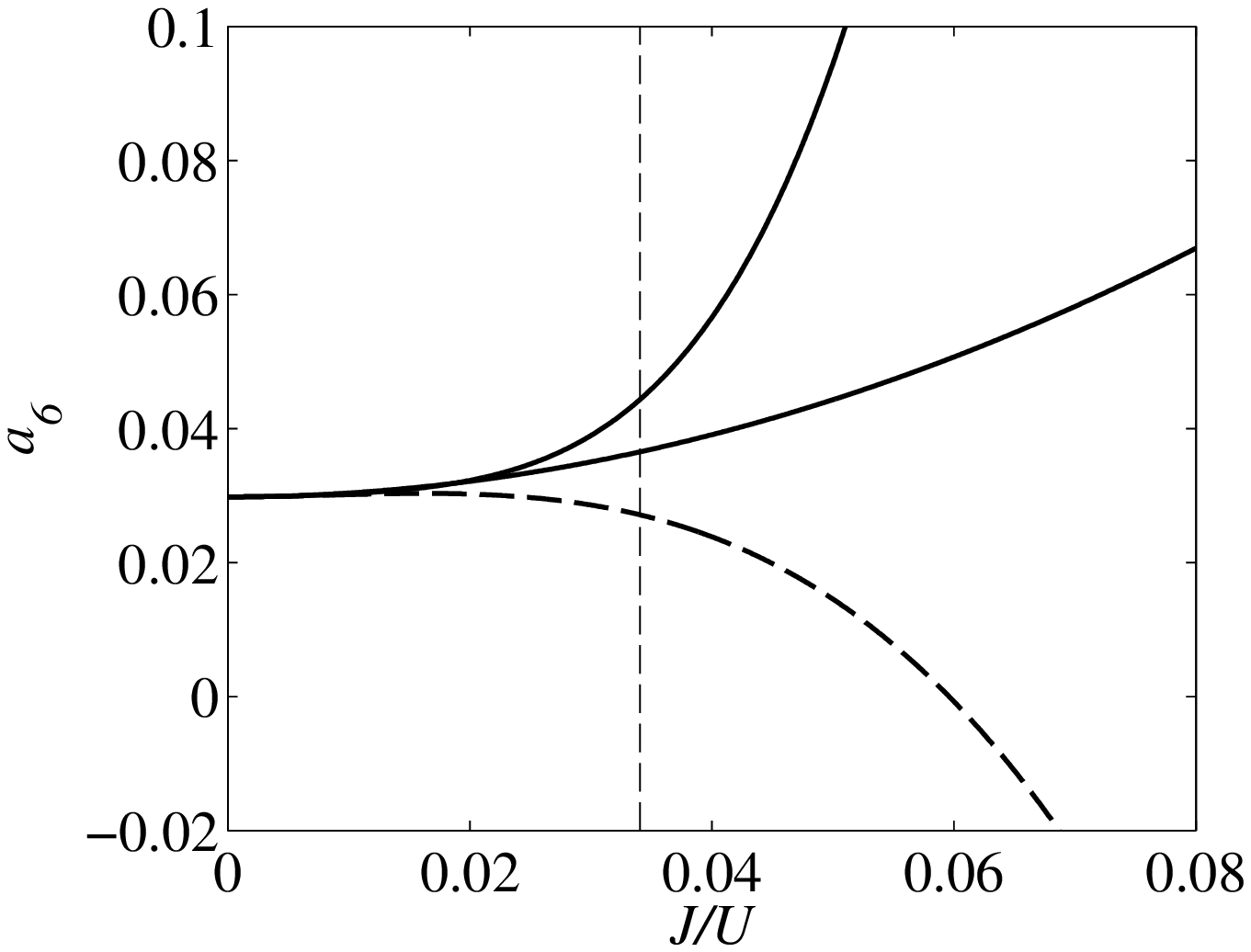}
\caption{Upper panel: Successive perturbational approximants to the Landau
	coefficient $a_2$ for the 3D Bose-Hubbard model with scaled chemical 
	potential $(\mu/U)_{\rm c} = 0.393$, as corresponding to the tip of 
	the Mott lobe with filling factor $g = 1$. Starting with the leftmost 
	straight line and proceeding rightwards, the respective maximum 
	hopping orders $\nu_{\rm m}$ are 3, 2, 5, 4, 7, 6.
	Middle panel: As above for the coefficient $a_4$. Starting with 
	the leftmost line crossing the lower margin, and proceeding 
	counter-clockwise along the margin, maximum hopping orders 
	$\nu_{\rm m}$ are 6, 4, 2, 3, 5. 
	Lower panel: As above for the coefficient $a_6$. Maximum hopping
	orders $\nu_{\rm m}$, assigned as above, are 3, 2, 4.
	Vertical dashed lines mark $(J/U)_{\rm c}$.}
\label{fig:6}
\end{figure}

Perturbative approximants to the higher effective-potential coefficients $a_4$ 
and $a_6$ for the 2D Bose-Hubbard model are displayed in Fig.~\ref{fig:5};
note that the computation of $a_4$ with $\nu_m = 8$, or that of $a_6$ with
$\nu_m = 6$, necessitates to evaluate the perturbation series even to 12th 
order. In marked contrast to $a_2$, now the successive ``approximations'' 
do not approach each other with increasing $\nu_m$ in the vicinity of 
$(J/U)_{\rm c}$, but rather appear to diverge strongly in an alternating 
manner; increasing accuracy with increasing $\nu_m$ is achieved only for
comparatively small $J/U$. Evidently we are dealing with asymptotic series;
in order to deduce the true behavior of both $a_4$ and $a_6$ close to the 
phase transition one needs to convert the divergent weak-coupling series 
into convergent strong-coupling expansions. Techniques for doing this 
do exist~\cite{JankeKleinert95}, but would require some {\em a priori\/} 
information on the functional form of the true $a_4$ and $a_6$. A similar 
pattern is also observed in Fig.~\ref{fig:6}, in which corresponding plots of 
$a_2$, $a_4$, and $a_6$ for the 3D system with $g = 1$ are grouped together:
While successive estimates of the zero of $a_2$ actually come closer 
to each other with increasing $\nu_m$, allowing one to determine 
$(J/U)_{\rm c} \approx 0.03407$ by extrapolation, successive approximants 
to $a_4$ and $a_6$ repel each other in the vicinity of $(J/U)_{\rm c}$,
although this divergence appears to be somewhat less violent here than for 
$d = 2$. Again, our above estimate of $(J/U)_{\rm c}$ compares very favorably
with the QMC result $(U/J)_{\rm c} = 29.34(2)$~\cite{CapogrossoSansoneEtAl07}.

\begin{figure}
\centering
\includegraphics[scale=0.5,angle=0]{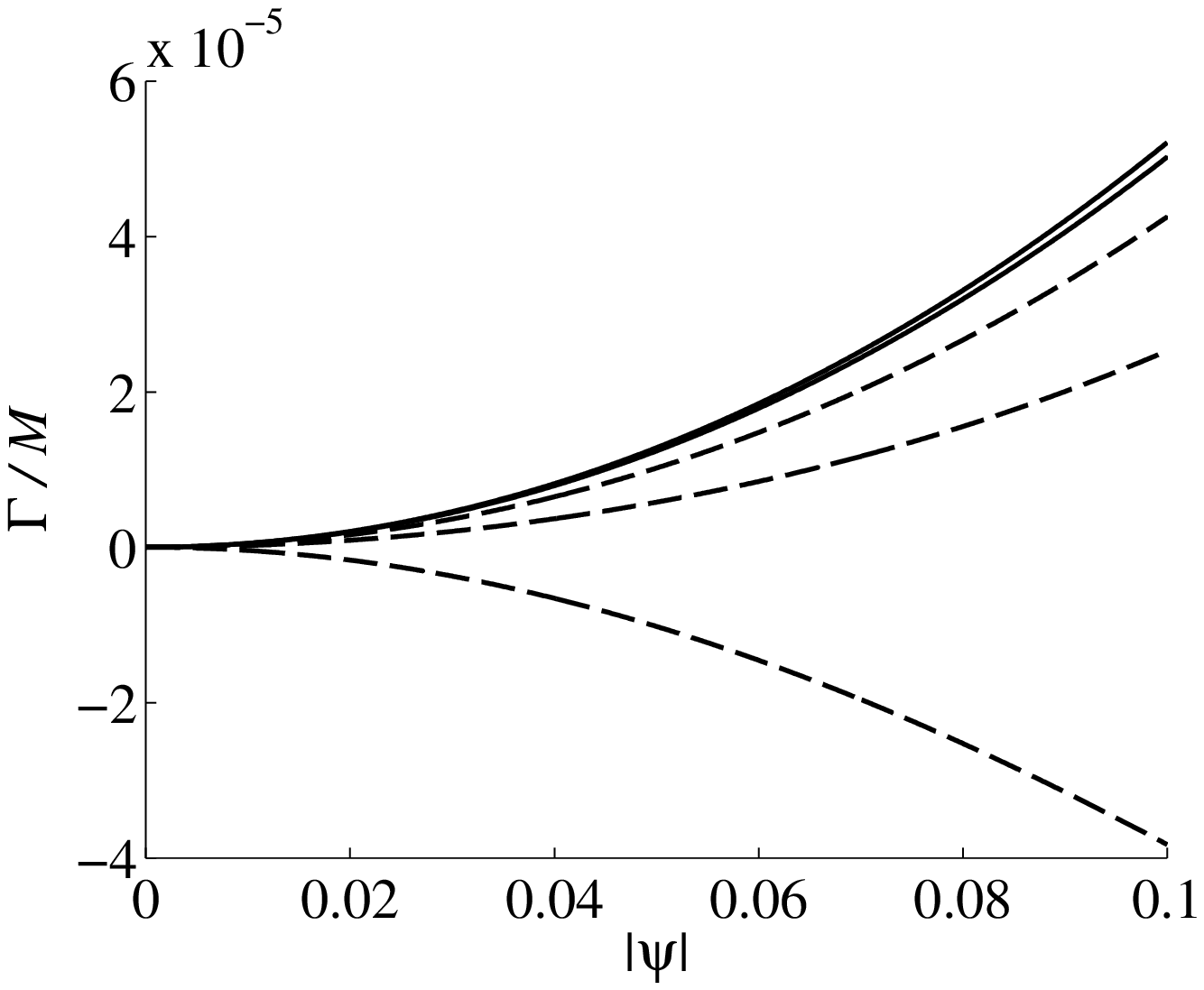}
\includegraphics[scale=0.5,angle=0]{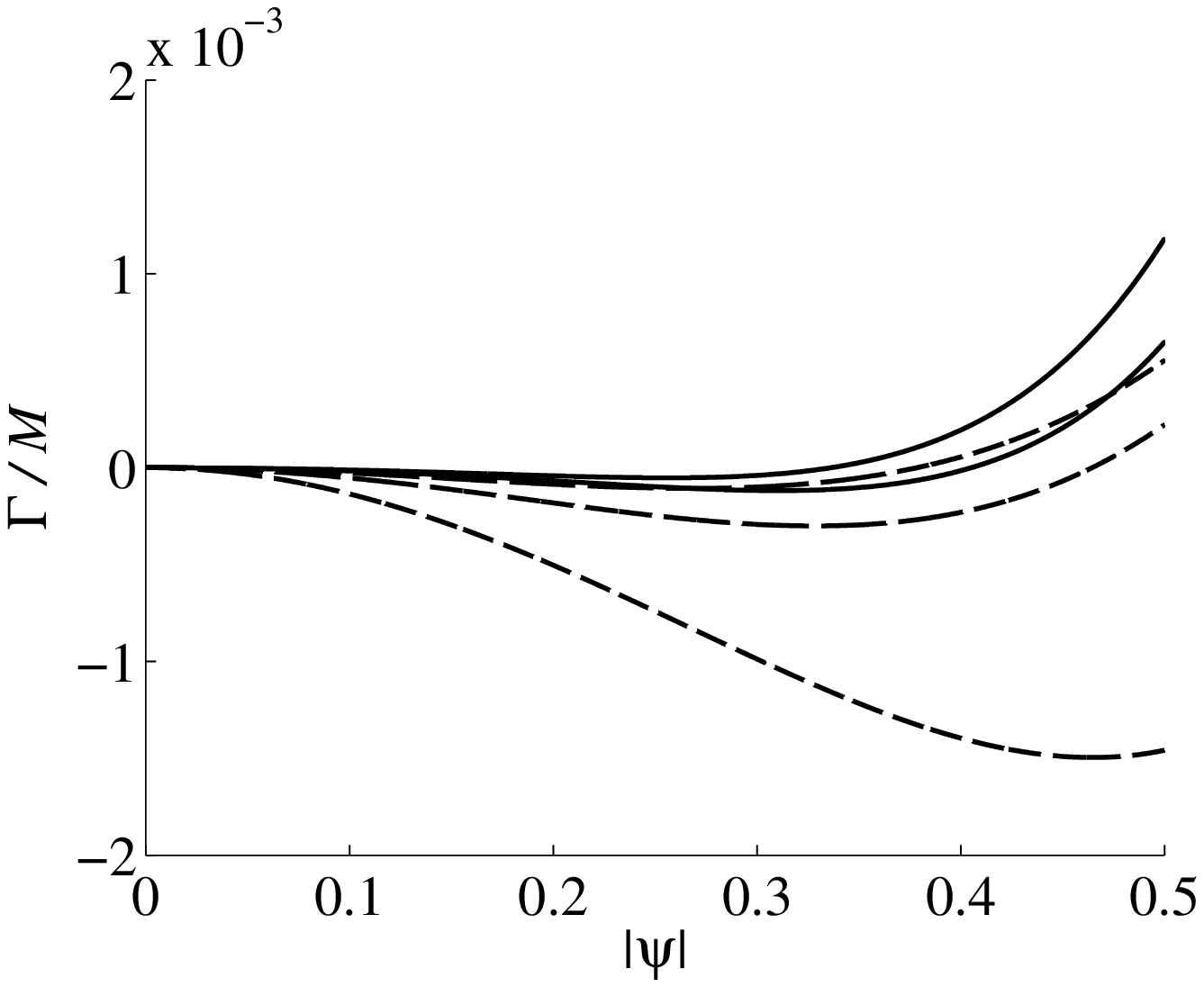}
\caption{Effective potential $\Gamma(\psi)/M$ evaluated for the 2D Bose-Hubbard
	 model with $(\mu/U)_{\rm c} = 0.373$, and $J/U = 0.055$ (above)
	 or $J/U = 0.059$ (below). Proceeding from bottom to top at the right
	 margin, maximum hopping orders $\nu_{\rm m}$ are 3, 5, 7, 4, 6 for
	 both panels.}
\label{fig:7}
\end{figure}

However, we are not primarily interested in the individual Landau 
coefficients~(\ref{eq:COP}), but rather in the full effective 
potential~(\ref{eq:EFP}). It is, therefore, interesting to observe that 
the divergent behavior of the perturbative approximants to $a_6$ appears to 
counteract the divergency of the approximants to $a_4$: Whereas the odd-order
approximants (dashed lines) appear to ``overshoot'' the true values of $a_4$ 
for both the 2D (Fig.~\ref{fig:5}) and the 3D system (Fig.~\ref{fig:6}), they
tend to ``undershoot'' the respective true values of $a_6$, and vice versa for
the even orders (full lines). Moreover, these higher coefficients enter into
$\Gamma$ only to higher orders in $|\psi|^2$, while we require accurate
know\-ledge of $\Gamma$ for small $|\psi|$. Thus, there is some hope that one 
still obtains a useful approximation to the effective potential even from the 
non-resummed coefficients. This hypothesis is supported by Fig.~\ref{fig:7}, 
which depicts successive approximants to the effective potential $\Gamma/M$ 
for the 2D system, as computed from $a_2$, $a_4$, and $a_6$ as functions of 
$|\psi|$. The upper panel refers to $J/U = 0.055$; the trend of the graphs
with increasing $\nu_{\rm m}$ suggests that the higher-order approximants 
indeed yield an acceptable estimate of $\Gamma$ in the full range 
$0 \leq |\psi| \leq 0.1$ considered. The lower panel shows a similar plot 
for $J/U = 0.059$, quite close to the critical value, where one still finds 
a fairly reasonable behavior of the approximants even up to $|\psi| = 0.5$.

\begin{figure}
\centering
\includegraphics[scale=0.5,angle=0]{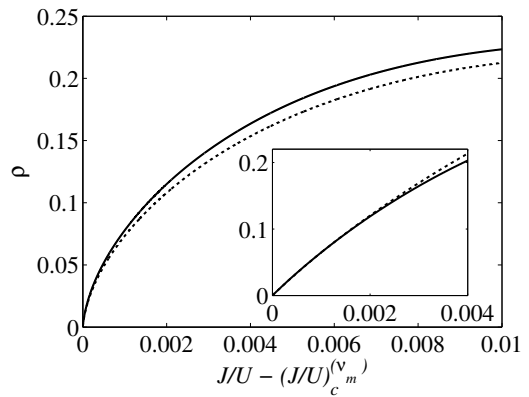}
\caption{Superfluid density $\varrho_{\rm s}$ (full lines) and condensate
	density $\varrho_{\rm c}$ (dotted) for $d = 2$ with $\nu_{\rm m} = 6$
	(main frame), and for $d = 3$ with $\nu_{\rm m} = 4$ (inset). While 
	the close-to-linear increase of both densities for $d = 3$ yields the 
	expected mean-field exponents $\beta_{\rm c} = \zeta = 1$, one finds 
	nontrivial exponents for $d = 2$. The superfluid densities have been 
	computed with the twist $\theta/\ell = 0.001$.}
\label{fig:8}
\end{figure}

This observation allows us to proceed, albeit tentatively, with the 
perturbative approximants to the coefficients~(\ref{eq:COP}), and to use 
these for computing the condensate density~$\varrho_{\rm c}$ by means of 
Eq.~(\ref{eq:CDN}). Here we admit even-order approximants only, since 
according to Figs.~\ref{fig:5} and \ref{fig:6} only even~$\nu_{\rm m}$ provide 
positive~$a_6$, and hence guarantee a stable, confining effective potential 
when terminating the Landau expansion~(\ref{eq:EFP}) after the sixth-order 
term; approximants with odd $\nu_{\rm m}$ are disregarded. Moreover, when 
Eq.~(\ref{eq:RSF}) is evaluated likewise with a sufficiently small value of 
the twist $\theta/\ell$, it yields a corresponding estimate of the superfluid 
density $\varrho_{\rm s}$. Figure~\ref{fig:8} shows results thus obtained with 
$\nu_{\rm m} = 6$ for $d = 2$ (main frame), and with $\nu_{\rm m} = 4$ for 
$d = 3$ (inset). Both densities initially increase about linearly for $d = 3$, 
heralding trivial (mean-field) critical exponents $\beta_{\rm c} = 1$ for 
$\varrho_{\rm c}$, and $\zeta = 1$ for $\varrho_{\rm s}$. This is to be 
expected, because the 3D Bose-Hubbard system belongs to the universality class 
of the 4D $XY$ model; since $d = 4$ is the upper critical dimension of this 
latter model, mean-field theory provides the correct critical exponents for 
this dimension, and all higher ones. On the other hand, the 2D Bose-Hubbard 
system falls into the 3D $XY$ universality class; in this case the exponents 
are nontrivial. Thus, although the Bose-Hubbard system with $d = 3$ spatial
dimensions is computationally more demanding, $d = 2$ is the case of main 
interest. Indeed, Fig.~\ref{fig:8} clearly indicates that the exponents for 
$d = 2$ must be significantly lower than $1$; from the fact that the 2D 
condensate density $\varrho_{\rm c}$ (dotted) lies below the superfluid 
density $\varrho_{\rm s}$ (full line) one deduces that the exponent 
$\beta_{\rm c}$ of $\varrho_{\rm c}$ is larger than the exponent $\zeta$ of 
$\varrho_{\rm s}$. This finding is in line with the Josephson 
relation~\cite{FisherEtAl73,RudnickJasnow77,Josephson66}
\begin{equation}
	\zeta = \beta_{\rm c} - \eta\nu \; ,
\end{equation}
where $\nu$ is the critical exponent of the correlation length, as already
referred to in the Introduction, and $\eta$ is the critical exponent of the
correlation function.

\begin{figure}
\centering
\includegraphics[scale=0.5,angle=0]{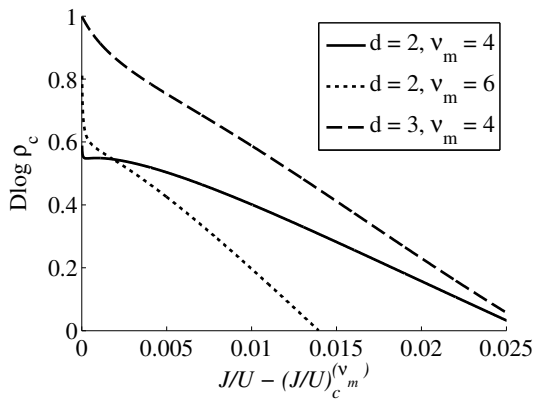}
\caption{Logarithmic derivative~(\ref{eq:LOG}) of the condensate density 
	$\varrho_{\rm c}$, computed according to Eq.~(\ref{eq:CDN}) for 
	$d = 2$ with both $\nu_m = 4$ and $\nu_m = 6$, and for $d = 3$ with 
	$\nu_m = 4$. Observe that continuing the linear part of the graph for 
	$d = 3$ to  $J/U - (J/U)_{\rm c} = 0$ yields $\beta_{\rm c} = 1$ with 
	reasonable accuracy, whereas the data for $d = 2$ clearly suggest a 
	smaller value.} 
\label{fig:9}
\end{figure}

Assuming now that the densities behave as 
\begin{equation}
	\varrho \propto \Big( J/U - (J/U)_{\rm c} \Big)^x
\end{equation} 
for $J/U$ somewhat larger than $(J/U)_{\rm c}$, the respective critical 
exponent $x$ is unveiled by computing the logarithmic derivative
\begin{equation}
	{\rm Dlog} \, \varrho = \frac{{\rm d} \log \varrho}
	{{\rm d} \log \big(J/U - (J/U)_{\rm c}\big)}
\label{eq:LOG}	
\end{equation}
and taking the limit 
\begin{equation} 
 	x = \lim_{J/U - (J/U)_{\rm c} \to 0} \; {\rm Dlog} \, \varrho \; . 
\label{eq:LIM}
\end{equation} 
In Fig.~\ref{fig:9} we plot the logarithmic derivative~(\ref{eq:LOG}) of
$\varrho_c$ for both $d = 2$ as obtained from approximations with either 
$\nu_{\rm m} = 4$ or $\nu_{\rm m} = 6$, and for $d = 3$ with $\nu_{\rm m} = 4$.
Evidently these derivatives behave almost linearly over wide ranges of $J/U$, 
with the exception of the immediate vicinity of $(J/U)_{\rm c}$. But this
latter regime has to be ignored anyway, because all our numerical results 
are given in terms of power series, thus isolating a single term close to
$(J/U)_{\rm c}$, whereas several powers have to combine in order to mimic 
non-integer exponents. Therefore, we obtain plausible finite-order estimates 
$\beta_{\rm c}^{(\nu_{\rm m})}$ of the condensate-density exponent 
$\beta_{\rm c}$ by extending the linear slopes to $J/U - (J/U)_{\rm c} = 0$: 
To begin with, for $d = 3$ we have $\beta_{\rm c}^{(4)} \approx 0.94$, quite 
close to the known exact value $\beta_{\rm c} = 1$. In view of our still shaky 
line of reasoning concerning the partial compensation of the divergencies 
plaguing the individual coefficients $a_4$ and $a_6$, this finding is quite 
encouraging.

Turning at last to the truly interesting case $d = 2$, and proceeding as 
above, we obtain the estimates $\beta_{\rm c}^{(4)}$ and $\beta_{\rm c}^{(6)}$ 
listed in Tab.~\ref{tab:1}; a linear fit of these data over $1/\nu_{\rm m}$ 
then provides the limit $\beta_{\rm c} = 0.7029$ for $\nu_{\rm m} \to \infty$. 
Similarly, we compute finite-order estimates $\zeta^{(\nu_{\rm m})}$ of 
the superfluid-density exponent $\zeta$, with an imposed twist of either 
$\theta/\ell = 0.01$, or $\theta/\ell = 0.001$. First the extrapolation to 
$\nu_{\rm m} = \infty$ is done separately for each twist, as is also documented
in Tab.~\ref{tab:1}; then a further linear extrapolation to $\theta/\ell = 0$ 
gives the final value $\zeta = 0.6681$.

\begin{table}
\caption{Finite-order estimates of the critical exponent~$\beta_{\rm c}$ 
	for the condensate density~$\varrho_{\rm c}$, and of the critical 
	exponent~$\zeta$ for the superfluid density~$\varrho_{\rm s}$, as
	obtained for the 2D Bose-Hubbard model. Also listed are	their 
	extrapolations to infinite order, performed linearly in 
	$1/\nu_{\rm m}$. In the case of $\zeta$ two values of the twist 
	$\theta/\ell$ are considered, providing data which are extrapolated 
	separately to $\nu_{\rm m} = \infty$; a further linear extrapolation 
	then yields the desired limit for $\theta/\ell \to 0$.}
\label{tab:1}
\begin{center}
\begin{tabular}{|l|c|c|c|}\hline\hline
	  	&   $\beta_{\rm c}^{(\nu_{\rm m})}$   &  
	\multicolumn{2}{c|}{$\zeta^{(\nu_{\rm m})}$}   	      \\ \hline   
	$\nu_{\rm m} \;\; \backslash \;\; \theta/\ell$ & 
	     		      - & 0.001		& 0.01	      \\ \hline
	4 	 & 0.5715 	& 0.6446 	& 0.6463      \\
	6 	 & 0.6153 	& 0.6525 	& 0.6541      \\
	$\infty$ & 0.7029	& 0.6683	& 0.6697      \\ \hline
	$\theta/\ell \to 0$ & - & \multicolumn{2}{c|}{0.6681} \\ \hline\hline
\end{tabular}
\end{center}
\end{table}

\section{Discussion and outlook}
\label{sec:5}

The concept of the effective potential $\Gamma$, borrowed from 
field theory~\cite{ZinnJustin02,KleinertSF01}, provides an immediate 
connection between quantum critical phenomena and Landau's theory of phase 
transitions~\cite{SantosPelster09,BradlynEtAl09}. Knowledge of the 
coefficent~$a_2$ appearing in the Landau expansion~(\ref{eq:EFP}) of $\Gamma$ 
allows one to locate the phase boundary; knowledge of the higher coefficients 
in the vicinity of that boundary enables one to also monitor the emergence of 
the order parameter $|\psi_0|$, and hence to determine the associated critical 
exponent $\beta$. In Sec.~\ref{sec:4} we have applied this scheme to the Mott 
insulator-to-superfluid transition shown by the Bose-Hubbard model, after 
having computed the Landau coefficients by high-order perturbation theory. 
In principle, the condensate density then is given by the familiar relation
\begin{equation}
	\varrho_{\rm c} = |\psi_0|^2 = -\frac{a_2}{2 a_4}
\label{eq:RC4}
\end{equation}
for hopping strengths $J/U$ slightly above the critical value, so that it 
should suffice to calculate $a_2$ and $a_4$ only. However, our perturbative 
approximants to these coefficients suffer from the divergency of the
weak-coupling perturbation series, so that the above Eq.~(\ref{eq:RC4}) 
can be exploited only if our approach is supplemented by a controlled 
procedure for converting a divergent weak-coupling series into a convergent 
strong-coupling expansion, as exemplified in Ref.~\cite{JankeKleinert95}.
While such a procedure would require some {\em a priori\/} information on the 
behavior of the true $a_4$, here we have followed a different route, relying 
on the observation that the divergent behavior of the $a_4$-approximants is 
counteracted by that of the approximants to $a_6$, as seen in Figs.~\ref{fig:5}
and \ref{fig:6}. Therefore, we keep the sixth-order term in the Landau 
expansion~(\ref{eq:EFP}) and replace Eq.~(\ref{eq:RC4}) for $\varrho_{\rm c}$ 
by its extended analog~(\ref{eq:CDN}); the same approximation to $\Gamma$ 
is employed when evaluating Eq.~(\ref{eq:RSF}) for the superfluid density 
$\varrho_{\rm s}$. The critical exponent $\beta = \beta_{\rm c}/2$ for the 
order parameter and the exponent $\zeta$ for the superfluid density determined 
in this manner for the 2D Bose-Hubbard model are juxtaposed in Tab.~\ref{tab:2}
to the corresponding best known estimates computed for the 3D $XY$ universality
class~\cite{CampostriniEtAl01}. In the case of $\zeta$ we have employed the 
hyperscaling relation $\zeta = (d-2)\nu$, which reduces to $\zeta = \nu$ for 
$d = 3$ and thus equates $\zeta$ with the critical exponent $\nu$ for the 
correlation length~\cite{FisherEtAl73,RudnickJasnow77}. While the accuracy of 
our results is difficult to specify, and certainly does not match that achieved 
in Ref.~\cite{CampostriniEtAl01}, the very fact that the numerical values 
coincide to better than $1\%$ constitutes an impressive manifestation of 
universality.

\begin{table}
\caption{Comparison of the critical exponents $\beta = \beta_{\rm c}/2$ 
	and $\zeta$ obtained in this work for the 2D Bose-Hubbard model 
	with data computed by Campostrini {\em et al.\/} for the 3D $XY$ 
	universality class~\cite{CampostriniEtAl01}. In the case of $\zeta$
	the relation $\zeta = \nu$ is utilized.}
\label{tab:2}
\begin{center}
\begin{tabular}{|c|c|c|}\hline\hline
		& this work & Ref.~\cite{CampostriniEtAl01}	\\ \hline
	$\beta$ & 0.3515    & 0.3485(2)		\\	\hline
	$\zeta$ & 0.6681    & 0.67155(27)	\\	\hline\hline
\end{tabular}
\end{center}
\end{table}

Yet, our findings still have to be regarded as preliminary. Subsequent 
steps to be taken now should involve a more systematic processing of the 
perturbative data, combined with an improved fitting procedure and a 
reliable error estimate, and it will be important to answer the question 
whether the encouraging first results reported here can be made more 
precise~\cite{HinrichsEtAl13}.

Still, physics is not about producing numbers, but about providing insight.
It is, therefore, quite striking to observe that the elemental 2D Bose-Hubbard
model actually provides the critical exponents of the lambda transition, 
and it might be interesting to pin down the ``carrier'' of this universality 
in terms of the process-chain diagrams involved in the computation of the 
Landau coefficients. Is there, perhaps, some simple property of these diagrams 
which clarifies why the 2D model differs so significantly from the 3D one?    

Of course, the ultimate test of universality will also require an experimental 
high-precision measurement of the critical exponents of the 2D Bose-Hubbard 
model, as realized with ultracold atoms in planar optical lattices. 
Besides the experiments referred to in the Introduction, recent studies 
aiming at the single-site addressability of ultracold atoms in optical 
lattices~\cite{WurtzEtAl09,GemelkeEtAl09,BakrEtAl10,ShersonEtAl10} hold a 
particularly high promise in this respect, since such techniques may allow one 
to directly measure spatial correlation functions, and thereby to determine the 
exponents $\nu$ and $\eta$.  In any case, with ultracold atoms now entering 
the field of critical phenomena, far-reaching further developments lie 
ahead.

\begin{acknowledgments}
This work was supported by the Deutsche Forschungsgemeinschaft (DFG) 
under grant No.\ HO~1771/5. Computer resources have been provided 
by the HERO cluster of the Universit\"at Oldenburg. A.P.\ gratefully
acknowledges a fellowship from the Hanse-Wissenschaftskolleg. 
\end{acknowledgments}

\end{document}